\documentclass[aps,showpacs,showkeys,twocolumn,floatfix]{revtex4}

\usepackage{graphicx}
\usepackage{epsfig} 
\usepackage{epstopdf}
\usepackage{color}
\usepackage{amsfonts}
\usepackage{amssymb,amsmath}
\usepackage{appendix}
\usepackage{enumerate}

\begin{document}

\title{Threshold $q$-voter model}

\author{Allan R. Vieira$^{1}$, Celia Anteneodo$^{1,2}$}

\address{
$^{1}$ Department of Physics, PUC-Rio, Rua Marqu\^es de S\~ao Vicente, 225,
             22451-900, Rio de Janeiro, Brazil \\
$^{2}$ National Institute of Science and Technology for Complex Systems, Brazil}


\begin{abstract}
\noindent
 
We introduce the threshold $q$-voter opinion dynamics 
where an agent, facing a  binary choice, can change its mind when at least 
$q_0$ amongst $q$ neighbors share the opposite opinion. 
Otherwise, the agent can still change its mind with a certain probability $\varepsilon$. 
This threshold  dynamics contemplates the possibility of persuasion 
by an influence group  even when there is not full agreement among its members. 
In fact, individuals can follow their peers 
not only when  there is unanimity ($q_0=q$) in the lobby group, as assumed in the 
$q$-voter model, but,  depending on the circumstances, also when there is simple majority ($q_0>q/2$), 
Byzantine consensus ($q_0>2q/3$),  or  any minimal number $q_0$ amongst $q$. 
This realistic threshold gives place to emerging collective states and phase transitions 
which are not observed in the standard $q$-voter. 
The threshold $q_0$, together with the stochasticity introduced by $\varepsilon$,  
yields a phenomenology that mimics as particular cases the $q$-voter 
with stochastic drivings such as nonconformity and independence. 
In particular, nonconsensus majority states are possible, as well as  mixed phases. 
Continuous and discontinuous phase transitions can occur, but also  
transitions from fluctuating phases into absorbing states.

\end{abstract}



\maketitle

\section{Introduction}

In recent years, statistical physicists have proposed models of 
opinion dynamics to answer  questions of current interest about public 
opinion formation such as the upraise of 
debate polarization, fragmentation or consensus~\cite{review,book1,book2}. 
In most of the models, individuals can adopt one of two opposite opinions. 
Although there are some real situations that require 
to contemplate several discrete~\cite{Chen2005a,Holme2006a,plurality}, 
or also continuous opinions~\cite{Deffuant2000a,Singh,ramos2015a,allan}, 
the binary case constitutes the minimal setting to study opinion dynamics.  
The dual situation of deciding between two alternatives of similar attractiveness 
is faced in different scenarios, for instance,   
being favorable or unfavorable to a given proposal,  
or buying one of two similar products that compete in a market. 
Perhaps, the simplest binary opinion dynamics is given by 
the voter model, where each agent can flip its opinion by imitation (or contagion) 
of a randomly chosen neighbor~\cite{review,voter1,voter2}. 
In fact, imitation is a common and practical strategy for decision making, that 
shapes our behavior early in life~\cite{imitation}.

Contagion can be pairwise~\cite{allan,Biswas,Crokidakis2014a, Crokidakis2012a} or involve group 
interactions~\cite{ramos2015a,Galam1986a,Galam1997a,Sznajd,Galam2002a,Krapivsky2003a,Shao2009a,q-voter}.  
It can be outflow or inflow, that is, 
from the individual towards the neighbors  or from the neighborhood towards the individual, respectively.
 In particular, Castellano et al. proposed an extension of the fundamental voter model,  
the nonlinear voter (or $q$-voter) model \cite{q-voter}, 
where agents with binary opinions change mind when $q$ randomly chosen neighbors share the opposite opinion. 
This is a paradigmatic example of inflow group interaction that motivated the study of 
the evolution to consensus \cite{q-voter-1d,q-voter-1D-exit,q-voter-1D-exit2}, 
as well as the definition of variants of the original $q$-voter model, by introducing a probability 
of behaving as anticonformist, as stubborn, or of making random decisions~\cite{q-voter-noises,q-voter-zealots}.

The voter models have been investigated mainly in complete graphs or regular lattices but also in more 
realistic heterogeneous networks~\cite{
1-voter-complex1,1-voter-complex2,uncorrelated,javarone,q-voter-complex1,mapping,duplex,deadlocksWS,MFnetworks}  
and co-evolving~\cite{voter-coevolving1,voter-coevolving2,voter-coevolving3,voter-coevolving4,q-voter-coevolving} networks.

Now, we introduce a generalization of the $q$-voter model, 
motivated by the consideration that $q_0$ (with $0\le q_0\le q$)  of $q$ neighbors 
may be enough to change the opinion of an agent. 
In fact, individuals take decisions not only when 
there is unanimity ($q_0=q$) in the group of influencers but,  depending on the circumstances, 
the Byzantine consensus ($q_0>2q/3$), the majority ($q_0>q/2$),  or simply a minimal number $q_0$ 
following an idea among $q$ contacts may suffice. 
We will show that this realistic generalization leads to the appearance of collective states 
which are inaccessible through the $q$-voter dynamics, which only allows 
equal fraction of both opinions or total consensus.

The algorithm that defines this threshold opinion dynamics is introduced in Sec.~\ref{sec:model}. 
The analytical approach based on a Fokker-Planck description is presented 
in Sec.~\ref{sec:theo}. 
Results from simulations together with analytical considerations are presented 
in Secs.~\ref{sec:time} and \ref{sec:infinite}  and the Appendices. 
A summary and final remarks can be read in Sec.~\ref{sec:final}.

\section{Opinion dynamics}
\label{sec:model}

Although in principle the dynamics could be played in an arbitrary network of contacts, 
we consider a fully connected network of $N$ individuals.  
The opinions can have one of two different values, that we denote $\oplus$ and $\ominus$. 
The opinion of each individual can change influenced by a group of size $q$, 
according to the following algorithm:

\begin{enumerate}[(i)]

\item
An agent $i$ is selected at random.

\item
Other $q$ agents are selected and their opinions  are observed. 
Actually, repetitions are allowed. 
The possibility of repetitions mimics the fact that an individual can interact 
with a contact more than once.

\item
If there are at least $q_0$ contrary opinions 
(where the threshold value is $0 \le q_0 \le q$), then, 
the opinion $o_i$ of agent $i$ is inverted, imitating the neighbors with contrary opinion.

\item
Otherwise, the opinion of agent $i$ can also change with a probability $\varepsilon$, 
except if all in the $q$-lobby share the same opinion of site $i$. 

\end{enumerate}

When $q_0=q$, one recovers  the nonlinear $q$-voter~\cite{q-voter}. 
In particular, $q_0=q=1$ leads to the original voter model. 

\begin{figure}[b]
\begin{center}
\includegraphics[scale=0.3,angle=-90]{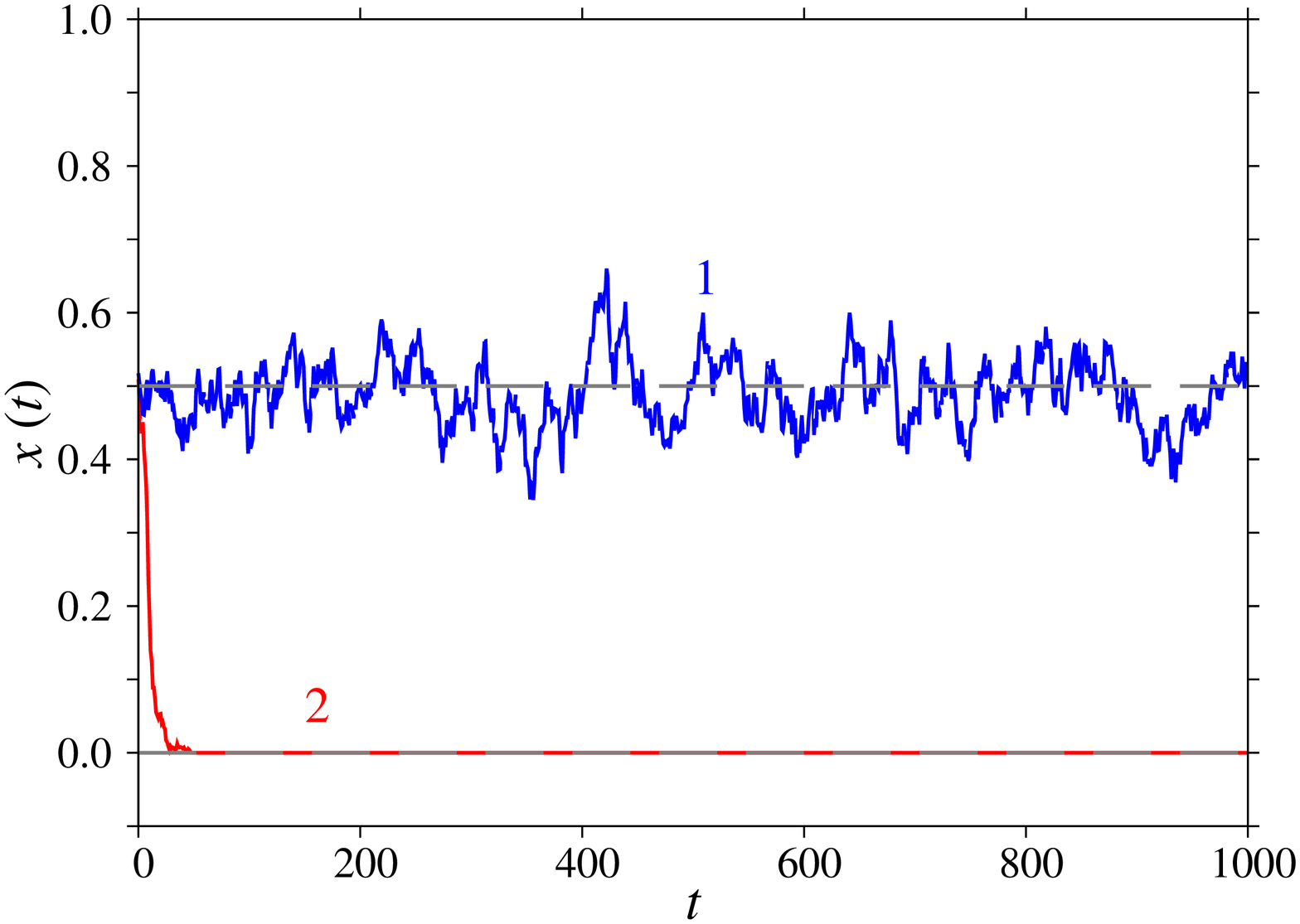}
\includegraphics[scale=0.3,angle=-90]{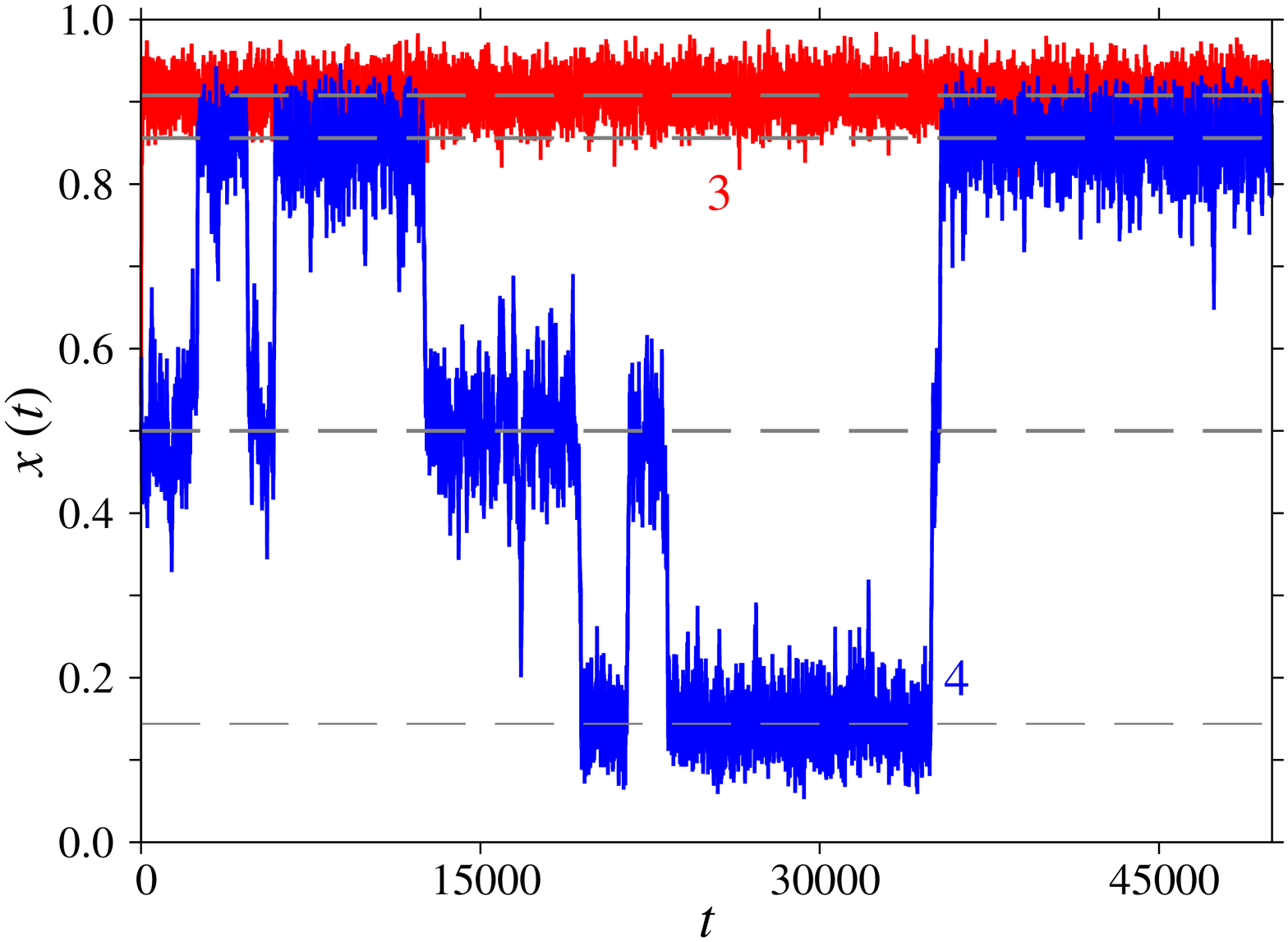}
\end{center}
\caption{Temporal evolution of the fraction of  $\oplus$  opinions $x(t)$. 
The parameters $(q,q_0,\varepsilon)$ are
1: (4,\,3,\,0.42), 2: (4,\,3\,,0.24), 3: (4,\,3\,,0.30), 4: (8,\,7\,,0.173).  
In all cases $N=1500$ and, in the initial state, there is equal number of 
agents with opposite opinions ($x(t=0)=1/2$). 
Dashed horizontal lines correspond to the minima of each potential $V(x)$ obtained by 
integration of Eq.~(\ref{eq:v}).
}
\label{fig:dynamics}
\end{figure}

Higher $q_0$  means individuals that are more hard  to be convinced or less volatile. 
Typically $q_0 > q/2$, 
meaning that a majority of neighbors is required to convince. 
The threshold $q_0$ can also be interpreted as the requirement of a critical 
mass of people sharing the opposite opinion to change mind. 
For completeness, we will analyze the model 
for any integer $0\le q_0 \le q$. 
In the extreme case $q_0=0$, the selected agent changes mind no matter the opinion of the neighbors, 
that is, the individual acts as totally volatile, or independent.

The collective state can be characterized by the fraction of $\oplus$ opinions, $x(t)$. 
This is the relevant quantity that defines the opinion than wins. 
When $x=1/2$, there is balance of opinions, that is, a tie. 
Otherwise, one of the opinions wins, particularly full consensus occurs when $x=0$ or 1.
Except if $q_0=0$,  when consensus is reached, the dynamics freezes at that collective state. 
That is, consensus is an absorbing state. 

The time evolution of the collective state $x(t)$, 
obtained by running computationally the algorithm defined above, 
is illustrated in Fig.~\ref{fig:dynamics}, 
for different values of the parameters $(q,q_0,\varepsilon)$. 
The unit of time $t$ corresponds to $N$ iterations of the algorithm. 
A diversity of behaviors emerges. 
The average opinion can fluctuate around zero (case 1), 
indicating polarization with equipartition of opposite opinions, technically a tie within 
finite-size fluctuations. 
The system can attain consensus (case 2). 
It can also evolve visiting non trivial collective states (cases 3 and 4), 
where one of the two opinions dominates, but without full consensus.

In order to understand the phenomenology of the dynamics, observed through 
numerical simulations, we will resort to a theoretical approach.

\section{Theoretical approach}
\label{sec:theo}

A master equation for this problem can be derived along 
the lines of Ref.~\cite{q-voter,gardiner}. 
In this binary scenario, there are two subpopulations with opposite opinions. 
Let us call $n$  the number  of individuals with opinion $\oplus$ at a given instant. 
At an iteration of the algorithm (i)-(iv) defined in Sec.~\ref{sec:model}, 
$n$ can change to $n^\prime$, 
with a certain rate that we call $p(n\to n^\prime)$. 
The only possibilities with non null rate are $n^\prime=n+ 1, n, n-1$, since 
at most one flip of opinion happens at a time.%

Moreover, we consider the probability of inversion $G(y)$, 
which is the probability that the opinion $o_i$ of the chosen agent flips,  
given that it belongs to the subpopulation with $yN$ individuals. 
That is,  if $y=n/N$, then $G(y)$ gives the probability of flip from 
$\oplus$ to $\ominus$; while if $y=1-n/N$, then $G(y)$ gives the probability 
of flip from $\ominus$ to $\oplus$. 

In terms of the probability of flip, the transition rates can be written as
\begin{eqnarray} \nonumber
p_{n\to n-1} & = & x \;G(x) \, , \\ \label{eq:rates}
p_{n\to n+1} & = & (1-x) \;G(1-x) \, , \\ \nonumber
p_{n \to n} & = & 1 - p_{n\to n+1} - p_{n\to n-1}\,,
\end{eqnarray}
where $x\equiv n/N$. 

According to the algorithm, the probability of flip is governed by 
the parameters $q$, $q_0$ and $\varepsilon$.
A flip of $o_i$ occurs when the number of disagreeing neighbors 
exceeds a threshold $q_0$, but also, with chance $\varepsilon$  otherwise. 
Therefore the probability of a flip can be expressed as
\begin{equation}\label{eq:gg}
G(y,q,q_0,\varepsilon) = g(y,q,q_0) + \varepsilon [1-g(y,q,q_0)-y^q]  \;,
\end{equation}
where $g(y,q,q_0)$ is the probability of having at least $q_0$ opposite 
opinions amongst $q$ randomly drawn ones, 
when agent $i$ is in the side with $yN$ individuals. 
This probability is given by the summation of the different favorable cases, i.e.,
\begin{equation}\label{eq:g}
g(y,q,q_0) = \sum_{i = q_0}^{q} {{q}\choose{i}} (1-y)^i  y^{q-i} \;. 
\end{equation}
Notice that, implicitly, repetitions are allowed, as defined in item (ii) of the 
algorithm. 
In Eq.~(\ref{eq:gg}), the term $y^q$ discounts the case where all agents in 
the influence group share the same opinion of agent $i$, 
as stated in item (iv) of the algorithm.

Once the  transition rates are completely defined, we write a master 
equation that in the continuum limit 
can be approximated by a backward Fokker-Planck equation, namely,
\begin{equation} \label{eq:FP}
\frac{\partial P}{dt'} = v(x') \frac{\partial P}{dx'} + \frac{1}{2} D(x') 
\frac{\partial^2 P}{dx'^2} \,, 
\end{equation} 
where $P=P(x,t|x',t')$, the drift is
\begin{equation}\label{eq:v}
v(x) = (1-x)G(1-x,q,q_0,\varepsilon) - xG(x,q,q_0,\varepsilon) , 
\end{equation}
and the diffusion coefficient is
\begin{equation}\label{eq:D}
D(x) = [ (1-x)G(1-x,q,q_0,\varepsilon) + xG(x,q,q_0,\varepsilon) ]/N \;.
\end{equation}
Integrating Eq.~(\ref{eq:v}), we obtain the potential $V(x)$, such that $v(x)=-dV/dx$, 
and we arbitrarily set the integration constant such that $V(0)=0$.

\begin{figure}[t]
\begin{center}
\includegraphics[scale=0.3,angle=-90]{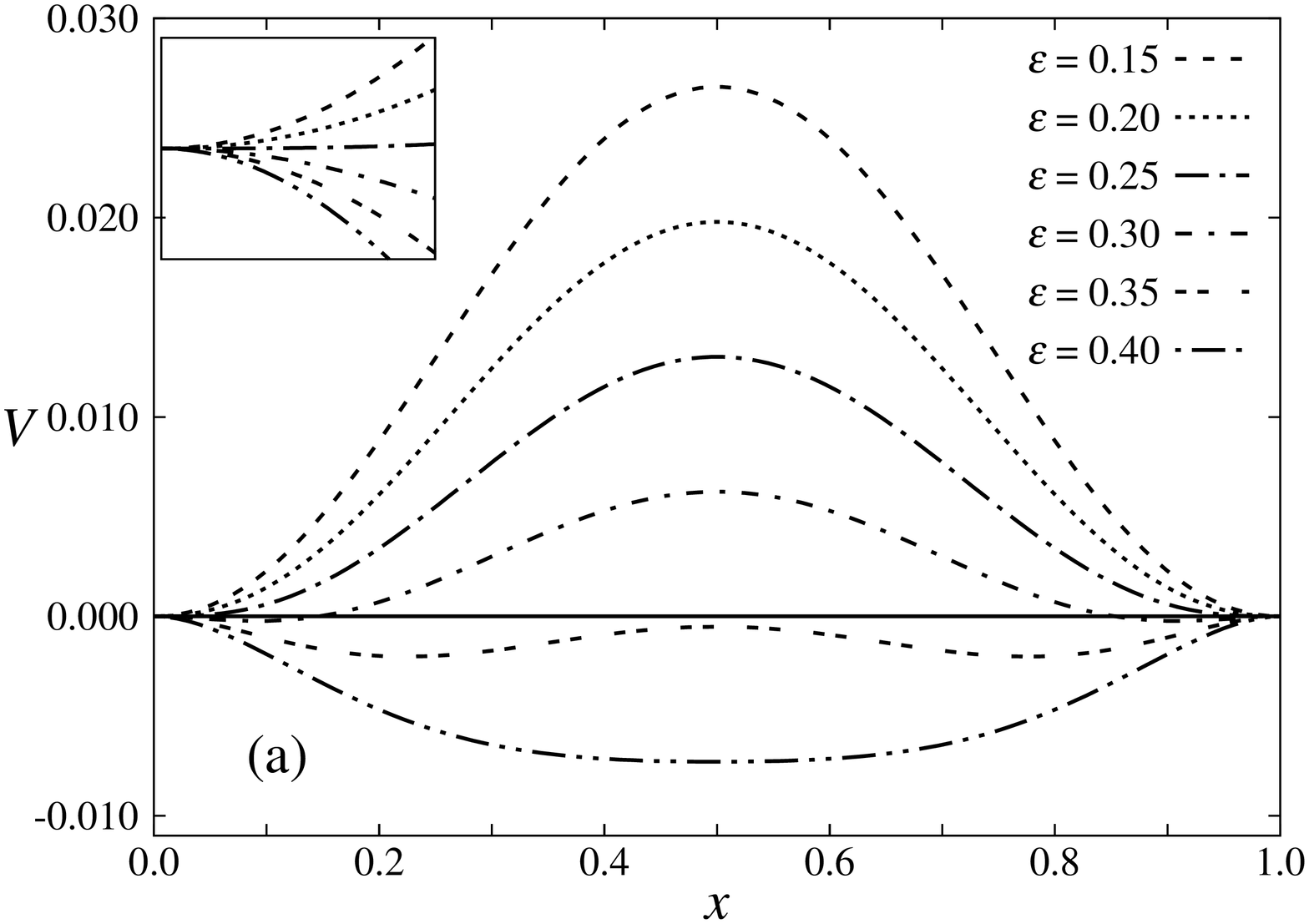}
\includegraphics[scale=0.3,angle=-90]{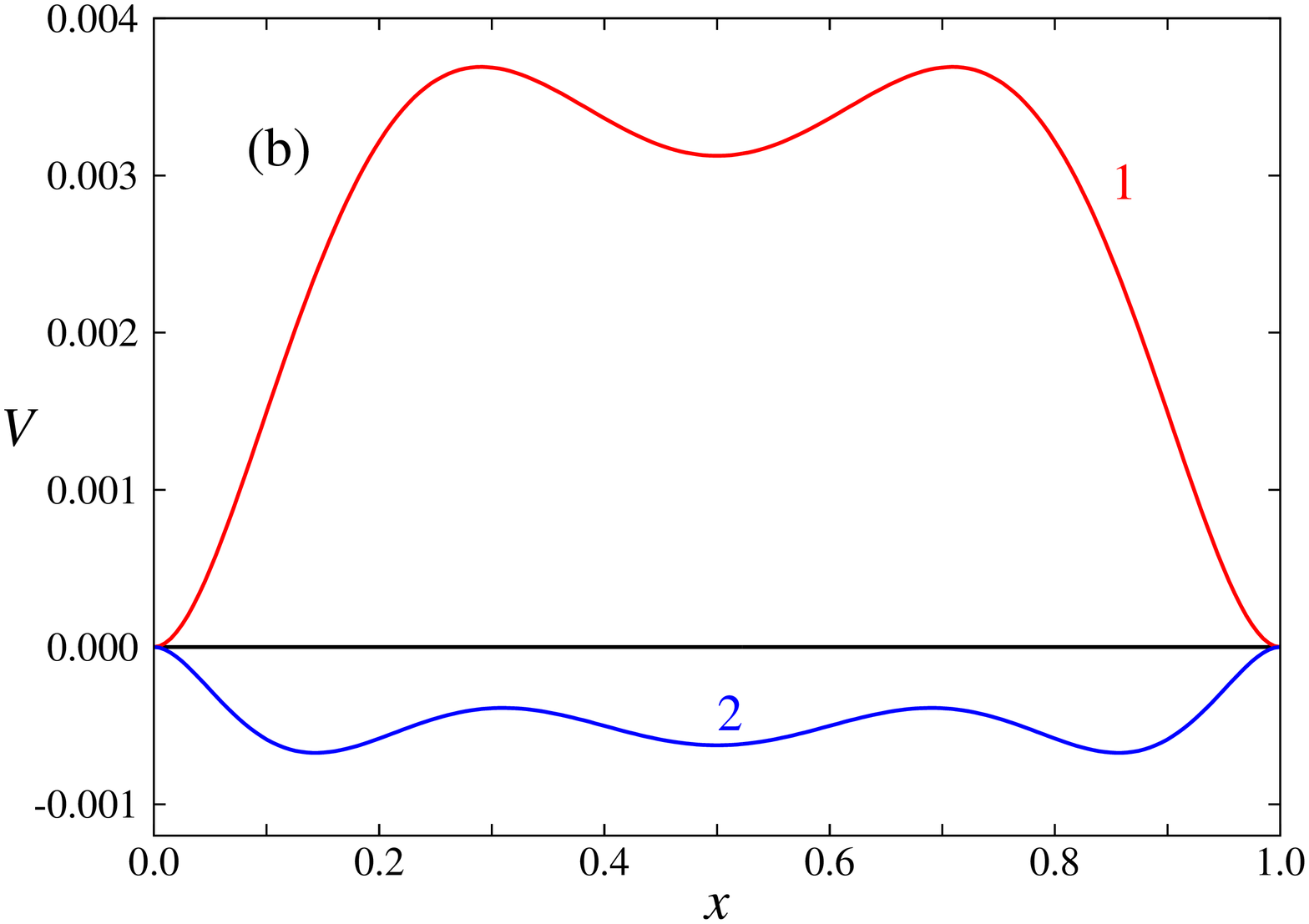}
\end{center}
\caption{Potential $V(x)$.  
(a) $q=4$ and $q_0=3$, for different values of $\varepsilon$ indicated on the figure. 
The inset is a zoom of the interval $x \in [0,0.01]$.
(b)  $(q,q_0,\varepsilon)=$  (4,\,2,\,0.12) and (8,\,7,\,0.173) for curves 1 and 2, 
respectively.  
Horizontal lines  at $V=0$ are drawn for comparison. 
}
\label{fig:V}
\end{figure}
 
Typical shapes of $V(x)$ are illustrated in Fig.~\ref{fig:V}, 
 for different values of the parameters. 
In all cases $V(x)$ has symmetry around $x=1/2$ as expected, due to the 
symmetric role of opposite opinions. 
The potential profile determines the probability with which the system  will dwell 
in each region of $0\le x \le 1$.

For sufficiently large  $\varepsilon$,   
the potential presents a single minimum at $x=1/2$,  
which leads to the formation of a disordered state fluctuating around equipartition of opinions, 
like in  case 1 of Fig.~\ref{fig:dynamics}. 
At a particular value $\varepsilon_1$,  the central minimum splits into two minima. 
This occurs when the second derivative of the potential $V(x)$ vanishes at the center. 
Using Eq.~(\ref{eq:v}), to set the equation $V''(x=1/2)=-v'(x=1/2)=0$, we obtain numerically   
$\varepsilon_1 \simeq 0.39$ in the case of Fig.~\ref{fig:V}(a). 
Below that value, the system can dwell in a collective state where one of the opinions dominates, 
without full consensus (see case 3 of Fig.~\ref{fig:dynamics}). 

Further decreasing $\varepsilon$, we find another particular value, $\varepsilon_2$, at which 
the second derivative of the potential $V(x)$ vanishes at the borders. 
From Eq.~(\ref{eq:v}), taking $V''(x=0)=-v'(x=0)=0$,  we obtain   
$\varepsilon_2=0.25$ in the case of Fig.~\ref{fig:V}(a). 
At this point, the extreme values at the borders become minima. 
Then, from any initial condition,  the system will directly evolve (without jumping potential barriers) 
towards an absorbing state of consensus, like in case 2 of Fig.~\ref{fig:dynamics}.  
 
Moreover, depending on the values of $q$ and $q_0$, there are still other possible 
shapes of the potential, different from those shown in Fig.~\ref{fig:V}(a). 
For instance, in  Fig.~\ref{fig:V}(b),  for $q=8$ and $q_0=7$, three local minima arise. 
Then, a trajectory like in case 4 of Fig.~\ref{fig:dynamics} can be observed.

However, in all these examples, at longer times than those shown in the figure, 
the system will eventually evolve to consensus, 
more rapidly the smaller the system. 
But, how long does it take to reach the absorbing state depending on the model 
parameters? 
How does it depend on the system size? 
These questions will be answered in the following section.

\section{Time to consensus}
\label{sec:time}

\begin{figure}[h]
\begin{center}
\includegraphics[scale=0.3,angle=-90]{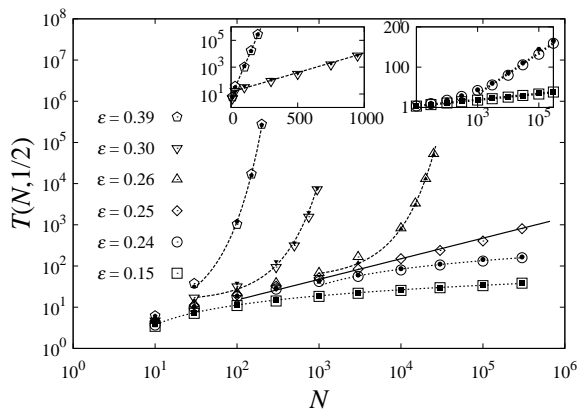}
\end{center}
\caption{Consensus time $T(N,x=1/2)$ vs. $N$, for $q=4$, $q_0=3$ and 
different values of $\varepsilon$ indicated on the figure. 
Hollow symbols correspond to the average over 200 samples, and 
full symbols to the value of the parameter that represents the mean value of 
an exponential (for $\varepsilon > 0.25$) or log-normal distribution (otherwise), 
obtained by a nonlinear curve fitting procedure (see Appendix~\ref{appendix1}). 
The solid line corresponds to $T\sim N^{1/2}$ for comparison.
The insets show the same data in log-linear and linear-log scales to exhibit the 
exponential and logarithmic growths, respectively.
}
\label{fig:T}
\end{figure}

We measured the average time, $T(N,1/2)$, that a system of size $N$ takes to reach the 
absorbing state of consensus,  when the initial condition is $x=1/2$ (equipartition of opinions). 
In Fig. \ref{fig:T}, we plot the numerical values  of $T$  as a function of $N$, 
choosing $(q,q_0)=(4,3)$ and different values of $\varepsilon$. 
The data points were obtained through two different procedures, 
with good agreement between them. 
The filled symbols represent the results obtained 
by performing the direct average over 200 samples. 
Alternatively, the hollow symbols were obtained from a fitting procedure, 
applied to the probability density function (PDF) of the time to consensus,  
as described in the Appendix~\ref{appendix1}.  
Succinctly, we plotted the mean value of the resulting PDF.

We observe that, in all cases, for finite $N$, it takes a finite time to reach consensus. 
Moreover, there is a critical value ($\varepsilon_c=\varepsilon_2= 0.25$, 
in the case of the example) for which 
the consensus time increases with system size as a power law, namely, $T\sim N^{1/2}$.  
This is the same critical behavior observed in the $q$-voter model~\cite{q-voter}. 
However, as we will see in the next section they correspond to different kind of transitions.
For $\varepsilon>\varepsilon_c$, in which case the potential presents at least one minimum 
localized outside the borders, the time increases exponentially. That is, following the Arrhenius 
law, since a potential barrier must be overcome to reach consensus. 
 Otherwise, when there are no potential barriers to jump,  
the evolution towards consensus occurs  exponentially fast (like in case 2 of Fig.~\ref{fig:dynamics}),  
in a time scale that in average increases logarithmically with $N$. 
 
In the next section, we will concentrate in the phases and transitions observed near the large size limit.

\section{Large-size systems}
\label{sec:infinite}

The smaller $N$,   the larger are the fluctuations of $x(t)$ and, 
the system can quickly reach the absorbing state of consensus, 
as can be observed in Fig.~\ref{fig:T}. Once consensus is reached, the system will 
be trapped there forever, unless an external perturbation takes the system out of that absorbing 
state.  

In the thermodynamic limit (TL),  fluctuations vanish, because according to Eq.~(\ref{eq:D}), 
$D(x) \sim 1/N$, then consensus is not reached in a finite time.  
Moreover, if a stable nonconsensus state (corresponding to the bottom of a potential well 
at $x\neq 0,1$) 
is accessible from a given initial condition, 
the system will evolve towards that steady state and remain there forever.

Visiting different states, spontaneously, requires fluctuations. 
Thus, trajectories like in  case 4 of Fig.~\ref{fig:dynamics} 
arise for sufficiently large but finite $N$ ($N=1500$ in the examples of the figure). 
These are the large-size systems whose evolution we will discuss in this section.

\begin{figure}[h]
\begin{center}
\includegraphics[scale=0.2,angle=180]{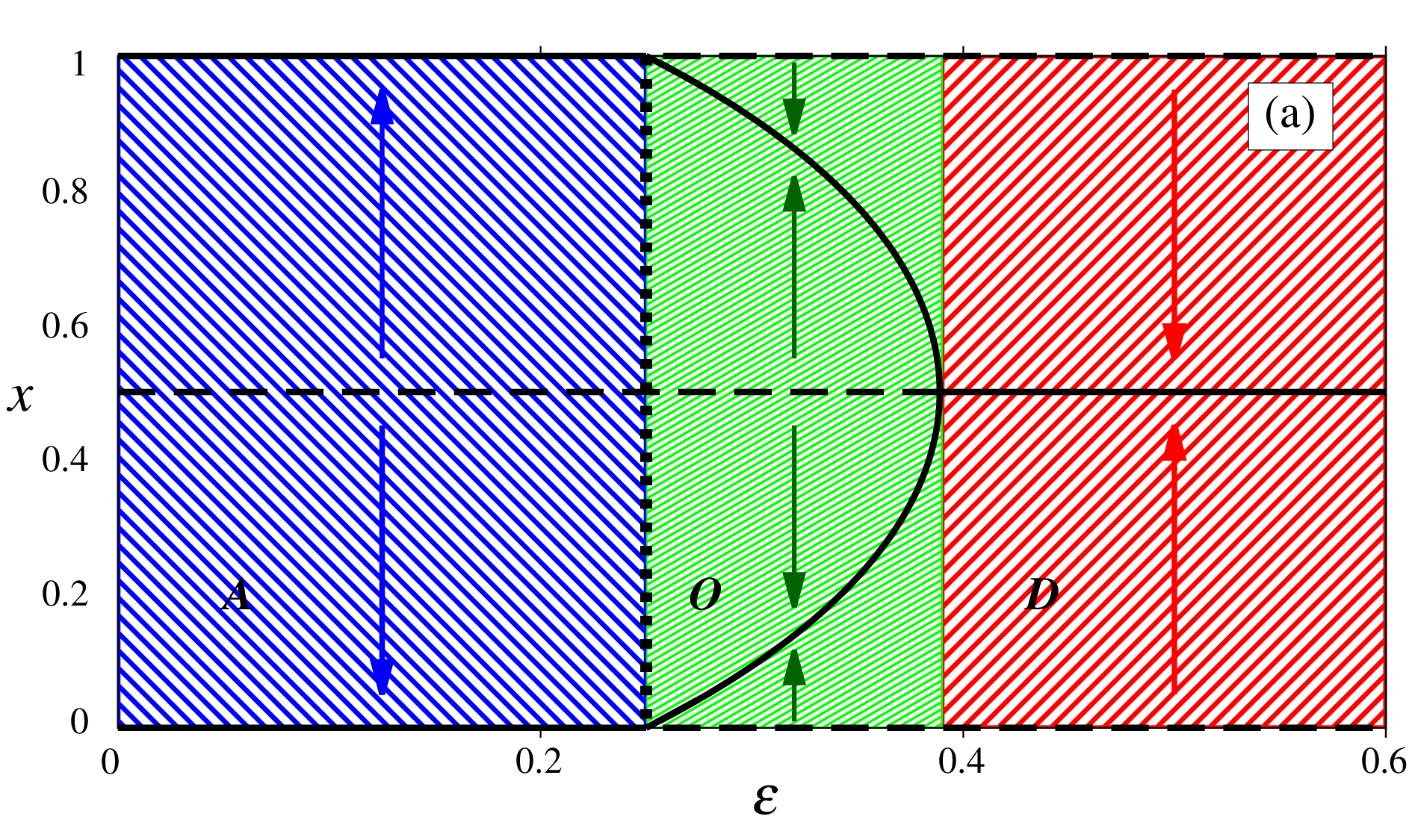}  \\[-2mm] 
\includegraphics[scale=0.2,angle=180]{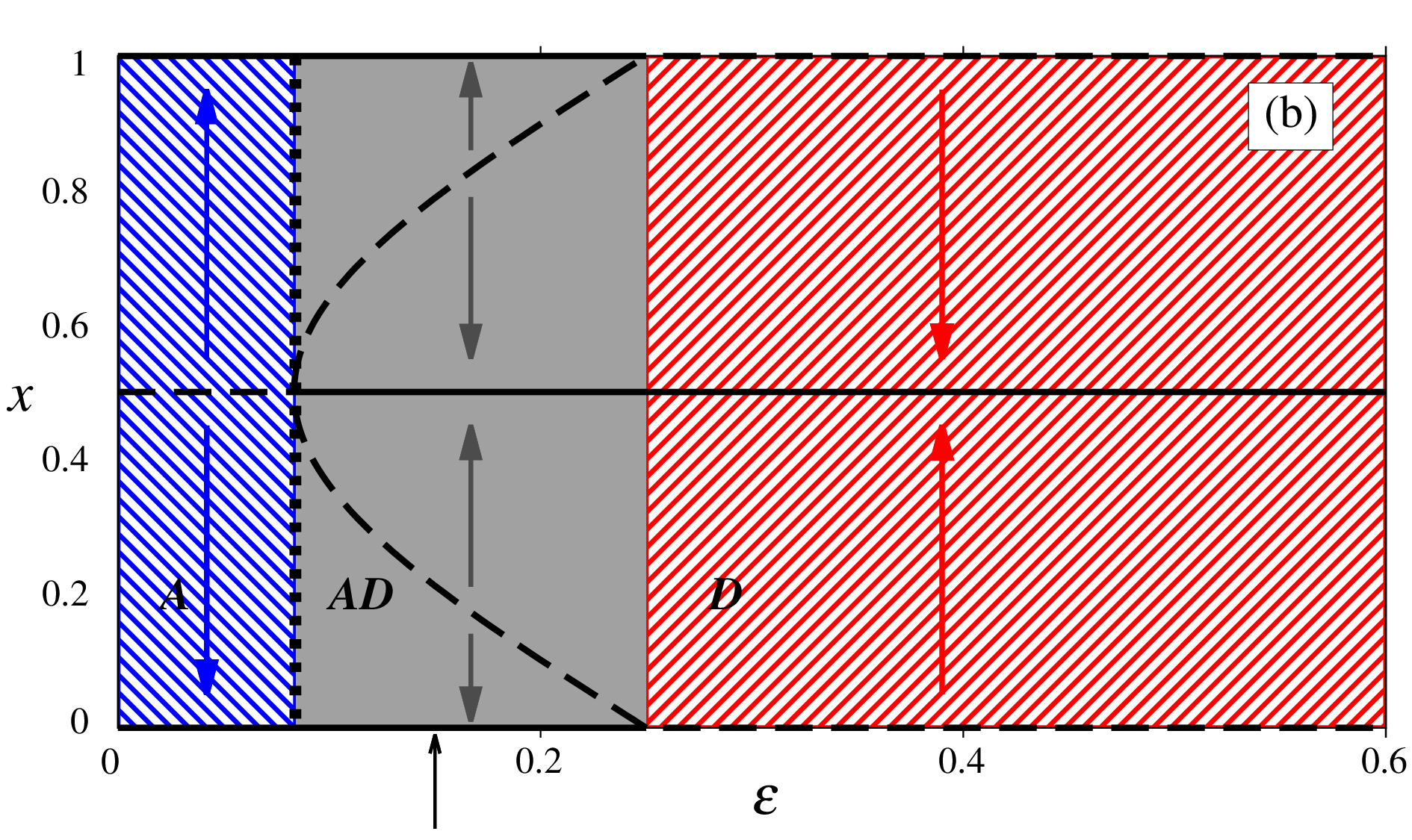}  \\[-2mm]   
\includegraphics[scale=0.2,angle=180]{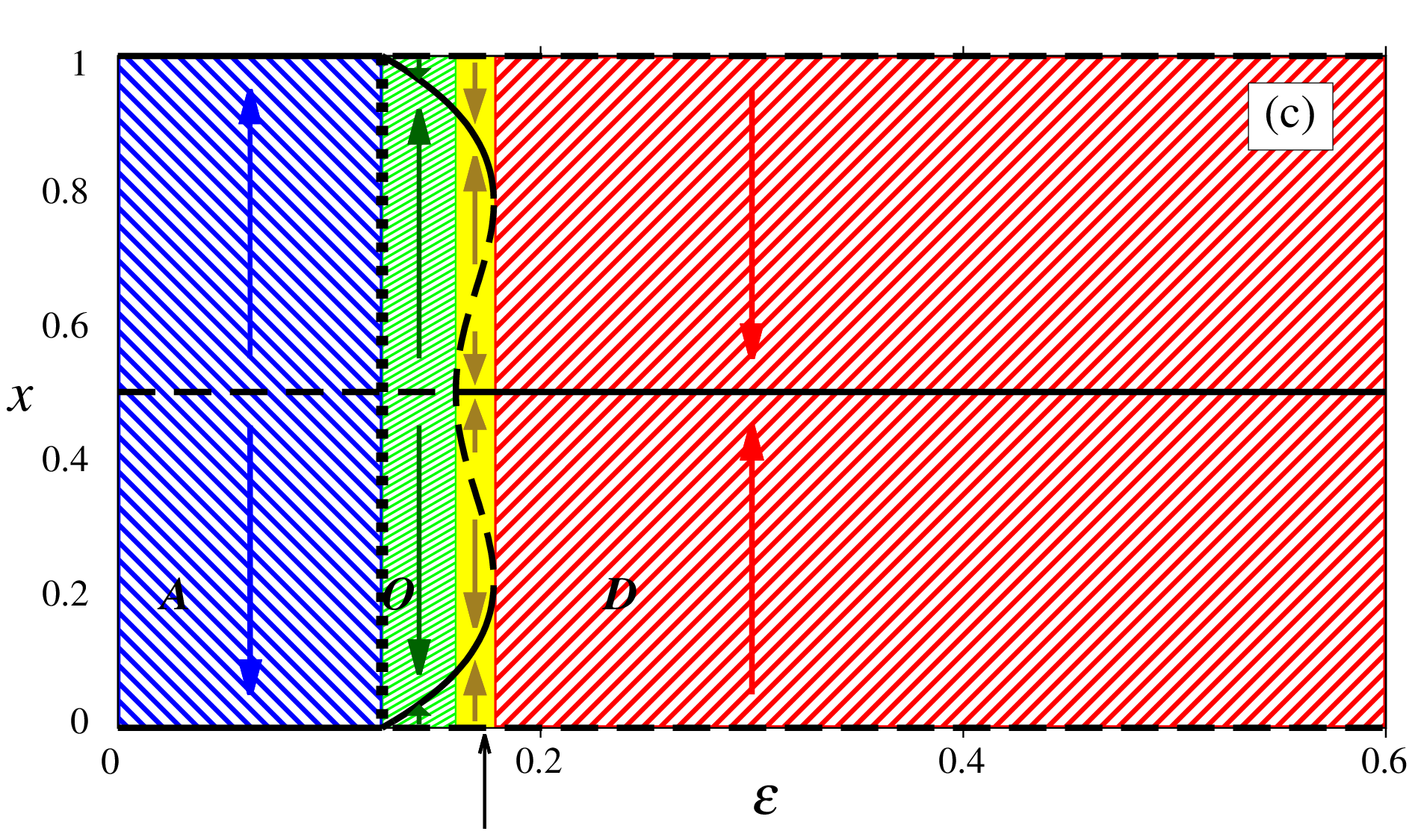}  \\[-2mm]    
\includegraphics[scale=0.2,angle=180]{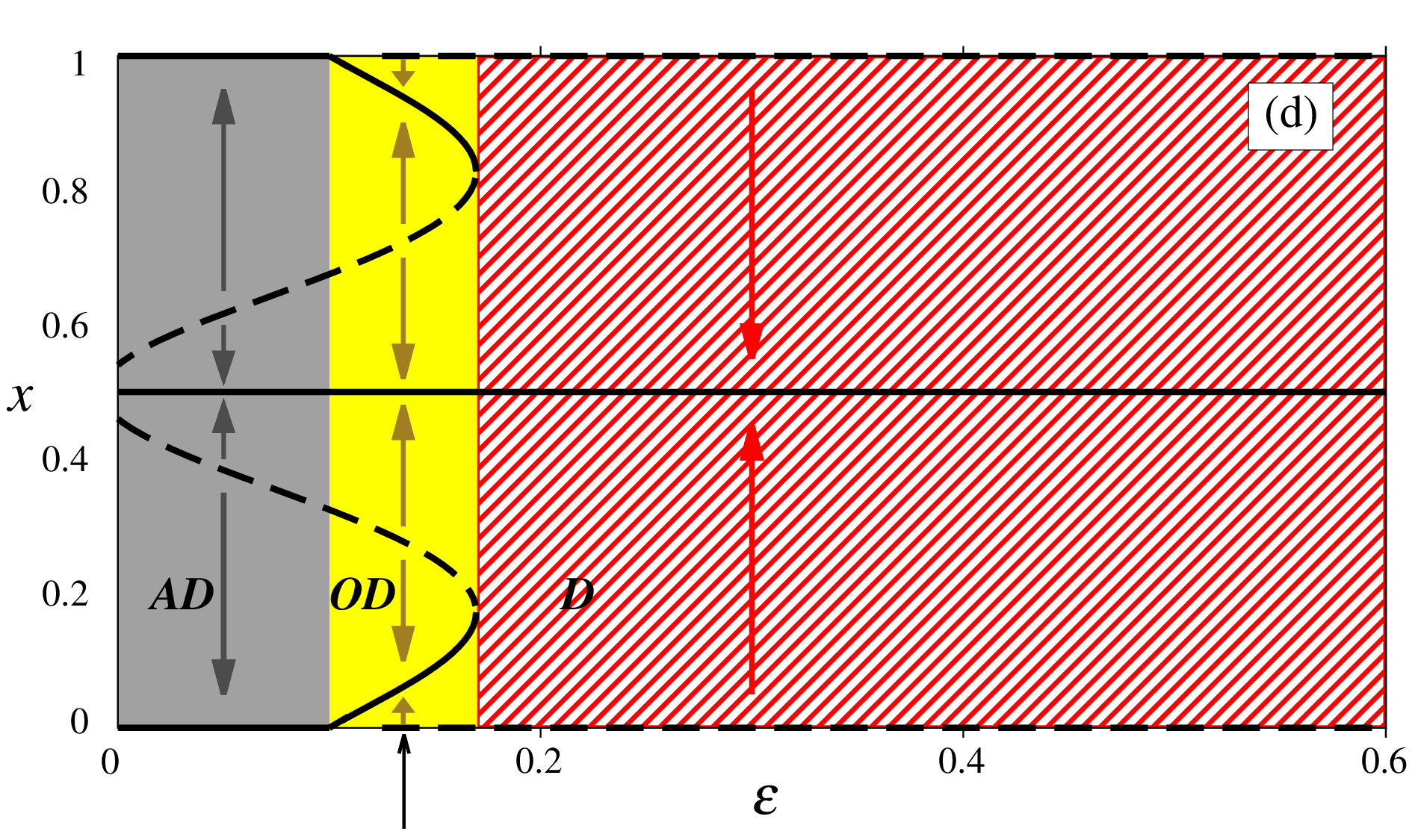}  \\[-10mm]   
\end{center}
\caption{Stability diagrams in the plane $x$-$\varepsilon$, 
for $(q,q_0)=$ (a) (4,3), (b) (4,2), (c) (8,7) and (d) (10,4).  
Solid (dashed) lines represent stable (unstable) values, 
corresponding to  minima (maxima) of potential $V$.
Each phase is represented by a different color and pattern: 
absorbing (A, blue), disordered (D, red), ordered (O, green), 
absorbing-disordered (AD, gray), ordered-disordered (OD, yellow).
The arrows inside the diagrams show the direction of the evolution, 
towards stable points, depending on the initial condition. 
The arrows below the abscissa axis show the values of $\varepsilon$ 
at which all the minima are equally deep. 
The vertical dotted lines indicate the transitions for which $T(N)$ 
presents critical behavior as in Fig.~\ref{fig:T}. 
}
\label{fig:stability}
\end{figure}

In Fig.~\ref{fig:stability} we show stability diagrams in the plane $\varepsilon$-$x$
for different values of $q$ and $q_0$, chosen to illustrate diverse phases and 
transitions between them. 
The black lines correspond to the extreme values of the potential, 
given by $v(x)=0$, for each value of $\varepsilon$. 
Solid (dashed) lines correspond to the minima (maxima) of the potential, 
meaning stability (instability) of steady states.  
It is noteworthy that consensus ($x=0$ or $x=1$) is an absorbing state in this model, 
then independently of whether it is a maximum or minimum of the potential, once 
the system reaches consensus, it can not escape.  
In that sense, consensus is always stable. 
However, if it is a maximum (minimum) of the potential, then it will be unstable (stable) 
with respect to external perturbations.

In  panel (a), notice that at small $\varepsilon$ the disordered state at $x=1/2$ is unstable while 
symmetric consensus states are stable. 
They loss stability beyond  $\varepsilon = 0.25$ where  two stable nonconsensus steady values emerge. 
From the viewpoint of a dynamical system, these transitions correspond to trans-critical bifurcations. 
The stable steady states collide at $\varepsilon\simeq 0.39$, where the disordered 
state  gains stability (super-critical pitchfork bifurcation). 
Three different phases can be identified:

``A'' (absorbing phase, in blue): 
the system tends to consensus from any initial condition. 
It is associated to a potential $V(x)$ that presents only minima located at the borders. 
 
``D'' (disordered phase, in red): opposite opinions are equilibrated within fluctuations, 
like a paramagnetic phase in magnetic systems. It is associated to a single 
central minimum of the potential, at $x=1/2$.

``O'' (ordered phase, in green):  there is unbalance of opinions, 
 like in a ferromagnetic phase. In this case, there is a majority or winner opinion. 
This phase is associated to the existence of two minima located at $x\neq 0,1/2,1$.
For the $q$-voter model (where $q_0=q$), it has been  argued \cite{q-voter} 
that an ordered phase (that the authors call Ising-ferromagnetic, directed-percolation-active) 
might also occur for values $2<q<3$. 
However, this possibility is virtual in that model as soon as $q$ is integer, 
while in the threshold voter, the ordered phase can be effectively 
observed for integer values of $q$ and $q_0$.

In panel (b),  there is stabilization of the disordered state at a critical value $\varepsilon=0.084$, 
in an inverse subcritical pitchfork bifurcation. At a second critical value $\varepsilon=0.25$, 
trans-critical bifurcations destabilize consensus states, 
while the disordered state remains stable. 
Between phases A and D, there is a mixed phase,
in the sense that both phases A and D can be observed depending on the initial condition. 

``AD''(absorbing-disordered, in gray):  emerges from a sub-critical pitchfork bifurcation. 
There is metastability. From the lower critical value of $\varepsilon$ 
the minima associated to consensus are deeper, at an intermediate value (indicated by a 
vertical arrow below the abscissa axis) the central minimum becomes deeper. 

In panel (c), between the phases O and D, we find another phase with metastability:

``OD'' (ordered-disordered, in yellow), characterized by a central minimum 
and two minima out of the borders. 
That is, either the two symmetric ordered states or the disordered one 
can be reached depending on the initial condition.  The stabilization of disorder 
occurs via an inverse subcritical pitchfork bifurcation, while the ordered states 
disappear due to  a saddle-node bifurcation.

Finally, in panel (d), we observe  transitions between phases AD and OD, which are trans-critical.

Notice that there are transitions involving absorbing states.  
Furthermore, between fluctuating phases, continuous and discontinuous transitions are possible. 
For instance, the transition between the phases O and D is continuous (see Fig.~\ref{fig:stability}a), 
while the transition between OD and D is discontinuous (see Fig.~\ref{fig:stability}d). 
In the latter case, hysteresis occurs, that is, the trajectories follow  
different paths when increasing and decreasing $\varepsilon$.  

We did not detect a transition between the phases AD and O, which would require 
a flat potential, like in the voter transition, 
nor a mixed phase of the type AO (absorbing-ordered).

\begin{figure}[h!]
\begin{center}
\includegraphics[scale=0.35,angle=-90]{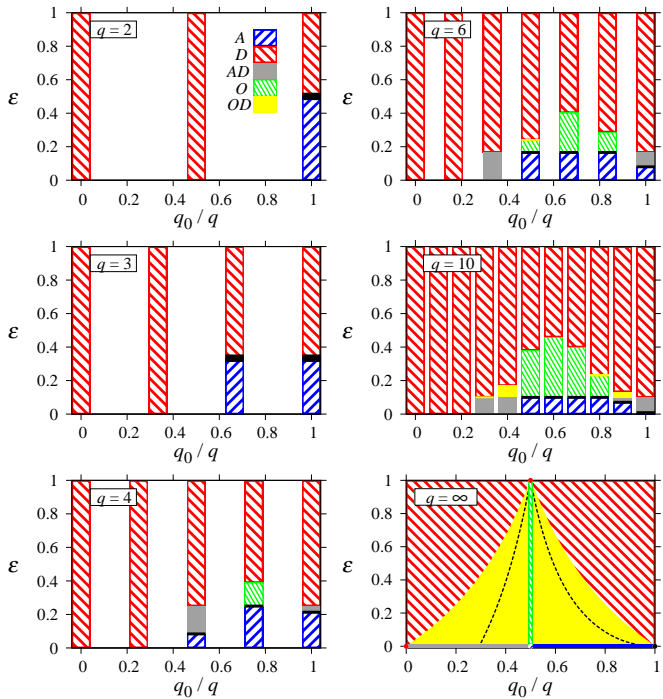}   
\end{center}
\caption{Phase diagram in the plane $(q_0/q)$-$\varepsilon$, for different values of $q$. 
Each phase is represented by a different color and pattern as in Fig.~\ref{fig:stability}. 
The black thick segments in cases $q=2,3$, represent the voter transition. 
Thin black segments indicate the critical point where $T\sim N^{1/2}$. 
Last panel (results obtained in Appendix~\ref{appendix2} for $q\to\infty$):   
the blue segment is where the potential is partially constant, and, 
at the dotted line, the three local minima of $V(x)$ possess the same value. 
}
\label{fig:diagrams}
\end{figure}

Phase diagrams in the plane $(q_0/q)$-$\varepsilon$, for all values of $q_0$, 
when $q=2,3,4,6,10$, are shown in Fig.~\ref{fig:diagrams}. We will discuss some special cases 
in the next subsection.

In the phase diagrams, the thin black segments indicate the 
critical point $\varepsilon_c$ where the consensus time increases as a power law with $N$, 
namely $T \sim N^{1/2}$, delimiting the regimes where the consensus 
time increases exponentially (above $\varepsilon_c$) or logarithmically (below $\varepsilon_c$) with $N$.
This critical behavior occurs between  A and AD phases in
the standard $q$-voter, while in this generalized version, 
the same critical behavior is additionally observed between phases A and O.

\subsection{Unanimity voter  ($q_0=q$)}

The $q$-voter, for which $q_0=q$, 
corresponds to the last column of each plot 
in Fig.~\ref{fig:diagrams}.  

For $q=2$ (top panel), when $\varepsilon$ increases, we observe the direct passage 
from  A to D, at  $\varepsilon = 1/2$. At that critical value, 
the potential is flat, $V(x)\equiv 0$, which corresponds to the voter transition, 
represented by a thick black segment. 
This transition also occurs for $q=3$ and $q_0=2,3$. 
In connection with the diagrams shown in Fig.~\ref{fig:stability}, 
this transition can be interpreted as the collapse of the intermediate 
regions O in panel (a) or AD in panel (b),   that disappear restoring the 
behavior with a single transition point.

For the higher lobby sizes $q=4, 6, 10$, 
we observe the appearance of the mixed phase AD (in gray) 
separating  A and D, like in panel (b) of Fig.~\ref{fig:stability}. 
This phase occurs for the interval $\varepsilon \in ( [q-1)]/[2^q-2],1/q)$. 
  Its boundaries can be derived from the conditions $v'(1/2)=0$ and $v'(0)=0$, 
respectively.
 
Therefore, this domain has maximal length at $q=8$ and 
shrinks completely when $q\to \infty$ (see Appendix~\ref{appendix2}). 
In that limit the voter transition is recovered (black circle).

For $q=1$ (voter model, not shown), the potential becomes 
flat, i.e., $V(x)\equiv 0$ for any $\varepsilon$.

\subsection{ Threshold voter $q_0\le q$}
\label{sec:zero}

When $q_0 < q$, the new phases O and OD, absent in the standard $q$-voter, emerge. 

The absorbing phase (A, blue)  shrinks when $q$ increases, 
surviving for progressively lower levels of the stochasticity introduced by $\varepsilon$.
This phase is absent for  $q_0/q <1/2$, due to the disordering effect of the 
independent behavior associated to small $q_0$.

The emergence of the ordered (O, green) phase requires a minimal lobby size $q\ge 4$. 
For increasing $q$, this phase becomes narrower and 
more central in the plane $(q_0/q)-\varepsilon$, such that 
in the limit $q\to \infty$, it occurs for any $0<\varepsilon<1$ at $q_0=q/2$ only. 
 
The order-disorder phase (OD, yellow) also appears for 
sufficiently large $q$ ($q\ge 6$). This domain grows for increasing $q$.

When  $q_0=q/2$ (central columns in the panels of Fig.~\ref{fig:diagrams} for even $q$), 
the ordered phase O emerges for $q\ge 6$ 
within the interval $\varepsilon \in (1/q,\varepsilon_{\rm max})$, 
where $\varepsilon_{\rm max}$ increases with $q$, such that the ordered phase 
occupies all the interval $0<\varepsilon<1$ in the limit $q\to\infty$.

The case $q_0=0$ was also included for completeness, although this case is trivial since the 
phase is disordered for any $\varepsilon$. This is because step (iii) of the algorithm is always 
satisfied, then a flip is certain, leading to disorder. Moreover, since step (iii) is always satisfied 
when $q_0=0$, then step (iv) is never  performed, hence the value of $\varepsilon$ is irrelevant.

\subsection{ Deterministic case $\varepsilon=0$}
\label{sec:large}

We start by discussing the large $q$ limit that admits simple analytical expressions. 
In the absence of the stochasticity introduced by $\varepsilon$, 
we find the AD phase (gray) for 
$q_0<q/2$  and the absorbing phase $A$ (blue) otherwise. 
But in this   blue horizontal segment, there is an hybrid situation: 
from Eqs. (\ref{eq:pot1}) and (\ref{eq:pot2}), 
the central region of the potential is flat
and the global minima are located at $x=0,1$. 
Only at $q_0 = q/2$ the flat region disappears.

At the extreme values we have always disorder when $q_0=0$, while in the opposite case
$q_0=q$, from Eq.~(\ref{eq:sign}),  $g(x)\simeq 0$ in the limit of large $q$ 
and consequently $G(x)$ also vanishes, 
generating a flat potential $V(x)\equiv 0$.

For intermediate values of $q_0$, the three regions of the potential $V(x)$ 
behave as follows (see Appendix~\ref{appendix2}: 
 the lateral regions  present their local minima 
at $x=\varepsilon/(1+\varepsilon)=0$, $x= 1/(1+\varepsilon)=1$. 
The intermediate region has length $ \left| 1 - 2q_0/q \right|$
and presents different shapes for   $ q_0 > q/2$ and $ q_0< q/2$, respectively:
 
$V(x) =  ( 1- q_0/q)^2/2$, i.e., it is constant.

$V(x) =  x(x-1) +( 1-  q_0/[2q])q_0/q$, which is a quadratic function 
with a minimum at $x=1/2$. 
 
Therefore, if $q_0\ge q/2$ consensus states are the only minima (phase A, blue), 
while, if $q_0<q/2$ there are three minima corresponding to consensus and disorder (phase AD, gray).

When $q$ is finite,  we notice in Fig.~\ref{fig:diagrams}, 
that the A region for $q_0 \ge q/2$ is maintained, except for $q \le 2$
while the AD  region shrinks as $q$ decreases 
giving place to the disordered phase $D$ (red) at low $q_0$.

\section{Summary and final remarks}
\label{sec:final}

 In this paper, we introduced the threshold $q$-voter, a model of opinion formation in a dual decision scenario~\cite{q-voter}, 
which is an extension of the $q$-voter~\cite{q-voter} and consequently also of the voter model~\cite{voter1,voter2}. 
The threshold $q_0$  is the minimal number, out of $q$ individuals forming an influence group, 
necessary to change the opinion of a randomly chosen agent. Like in the original version of the $q$-voter model, 
the agent can still change its mind with a probability $\varepsilon$, when the threshold condition is not satisfied.

We presented a characterization of the dynamics in the case of a fully-connected network. 
The threshold $q$-voter model exhibits a richer phenomenology than the 
standard $q$-voter, with new emergent phases, as can be observed in the phase diagram of Fig.~\ref{fig:diagrams}. 
The disordered (D) and  absorbing (A) phases 
that correspond to balance of opinions and consensus, respectively, and 
the mixed phase AD, where both collective states can occur depending on the initial condition, 
can be observed in the  standard $q$-voter. 
In contrast, the threshold effect can drive the collective state towards 
a nonconsensus ordered phase O, with a winner opinion (case 4 of Fig.~\ref{fig:dynamics}).  
This phase is not present in the $q$-voter model for the possible (integer) values of $q$.
Moreover, a new mixed phase (OD) also emerges (case 3 of Fig.~\ref{fig:dynamics}).

There are transitions involving absorbing states as in the $q$-voter, as well as new transitions between fluctuating phases. 
These can be continuous (like O-D) or discontinuous (like OD-D and OD-O) when changing parameter $\varepsilon$. 
 
Beyond the original version of the $q$-voter model, variants have been introduced \cite{q-voter-noises} 
by setting $\varepsilon = 0$ and adding stochastic drivings: 
one representing anticonformism (where agents can disagree from the influence group with certain probability) and 
another representing independence (where agents can take a random decision with a certain probability). 
These drivings prevent consensus from being an absorbing state. 
This is in contrast to the threshold $q$-voter, where for any value of $q$ and $q_0$,  
a finite-size system reaches consensus in a finite time, as illustrated in Fig.~\ref{fig:T}. 
That is, the stochasticity introduced by $\varepsilon$ hampers order but 
it does not prevent consensus from being an absorbing state, while the occurrence of 
independent and anticonformist decisions can take the system out from consensus due to random flips. 
Also notice that, although flipping by $\varepsilon$ can be associated to  
random decision making independently of the neighbors norm,  it occurs in step (iv) of the algorithm, 
only if the threshold condition is not satisfied. 
Therefore, in particular, when there is consensus there can not be flips, except if $q_0=0$, 
hence consensus is an absorbing state.
When $q_0=0$, step (iii) of the algorithm is always satisfied, then flipping is certain, 
leading to a disordered state,
and step (iv) is never executed, independently of $\varepsilon$. 
Since the chosen agent changes mind no matter the opinion of the
neighbors, it acts  independently  of the state of the  influence group. 
Although attenuated, this effect still occurs as $q_0$ increases up to $q/2$.

Certain analogies can be established between 
the threshold $q$-voter with $\varepsilon\neq 0$ 
and the variants with independence and anticonformism as defined above. 
For certain values of the flipping probability, both produce the ordered phase O, 
while the mixed phase OD is observed in the case of independence.  
 
Enhancing independence leads to the transitions  
O-OD-D \cite{q-voter-noises}  
(green-yellow-red regions in Fig.~\ref{fig:diagrams}). 
Looking at Fig.~\ref{fig:diagrams}, one sees that this sequence occurs for 
sufficiently large $q$, for instance, in the case $q=10$, setting $\varepsilon \simeq 0.1$ 
and decreasing $q_0$, which in fact we have associated to increasing level of independence.
The sequence also occurs by increasing $q_0$, from central values. 
In fact, if $q_0$ is large, it is difficult to satisfy the threshold condition and have a flip in step (iii) of the 
algorithm, because an agent needs to see almost unanimity in its neighborhood 
to change its mind. But then the algorithm proceeds to step (iv) and  
a flip in this step also represents a random change of opinion, independent of 
the dominant opinion in the influence group.

Increasing anticonformity  promotes the transition  O-D~\cite{q-voter-noises} (green-red in Fig.~\ref{fig:diagrams}).
In our model, this transition occurs typically for not too large $q$. 
For instance by starting from central values of $q_0$ and increasing $\varepsilon$. 
When $q_0$ is central, step (iii) of the algorithm is not implemented when there is a minority with opposite opinion, 
then a flip can occur in step (iv) leading to share the opinion of the minority, mimicking 
anticonformity or contrarian  decisions.  
  This effect increases with $\varepsilon$, then this parameter emulates the probability of the 
	anticonformist  driving.

Therefore,  the interplay between the threshold $q_0$ and the 
stochastic effect controlled by  $\varepsilon$ yields as particular cases the 
phenomenologies observed under the inclusion of independence or  anticonformity.

Let us also remark that consensus is always reached as a final state in the present model. 
However, while for $\varepsilon$ below a critical value $\varepsilon_c$ 
the time to consensus increases logarithmically with system size $N$, 
above $\varepsilon_c$, the growth is exponential. 
As a consequence,  a large system  in a mixed phase can visit metastable states before reaching consensus.
The emergence of metastability is particularly important in this context, 
because if the global opinion of a population can spontaneously change from one metastable state to 
another, then the result of the election will depend on the precise moment it occurs, 
differently to the case where there is a unique steady state. 
Moreover, a perturbation, like propaganda might easily switch the state of 
the system, redefining the result of the election. 
   
Finally, let us comment that the original voter, $q$-voter and their variants have been been studied not only in complete graphs and 
regular lattices, 
but also in diverse more realistic heterogeneous networks, such as Watts-Strogatz, Barabasi-Albert, random regular or even 
real networks (for the definition of these networks see for instance ~\cite{networks}). 
For the voter model, it was shown how network structure affects the $N$-dependence 
of the consensus time~\cite{1-voter-complex1,1-voter-complex2,uncorrelated}. 
For the $q$-voter, it was also shown that   the consensus reaching process is facilitated by properties such as
 long-range interactions~\cite{mapping}, randomness and density of a network~\cite{deadlocksWS,MFnetworks}. 
Moreover structure can also change the nature of phase transitions, like in duplex clique graphs~\cite{duplex}.
Therefore, as a perspective of future work,  it would be  interesting to analyze the  role of a topology 
in the threshold $q$-voter dynamics. Also interesting would be the study in co-evolving networks extending previous 
results for voter models~\cite{voter-coevolving1,voter-coevolving2,voter-coevolving3,voter-coevolving4,q-voter-coevolving} 
where in addition to consensus and coexistence, fragmentation of the network can emerge depending on the rewiring probability.

\appendix

\section{Consensus time distribution}
\label{appendix1}

\begin{figure}[h]
\begin{center}
\includegraphics[scale=0.25,angle=-90]{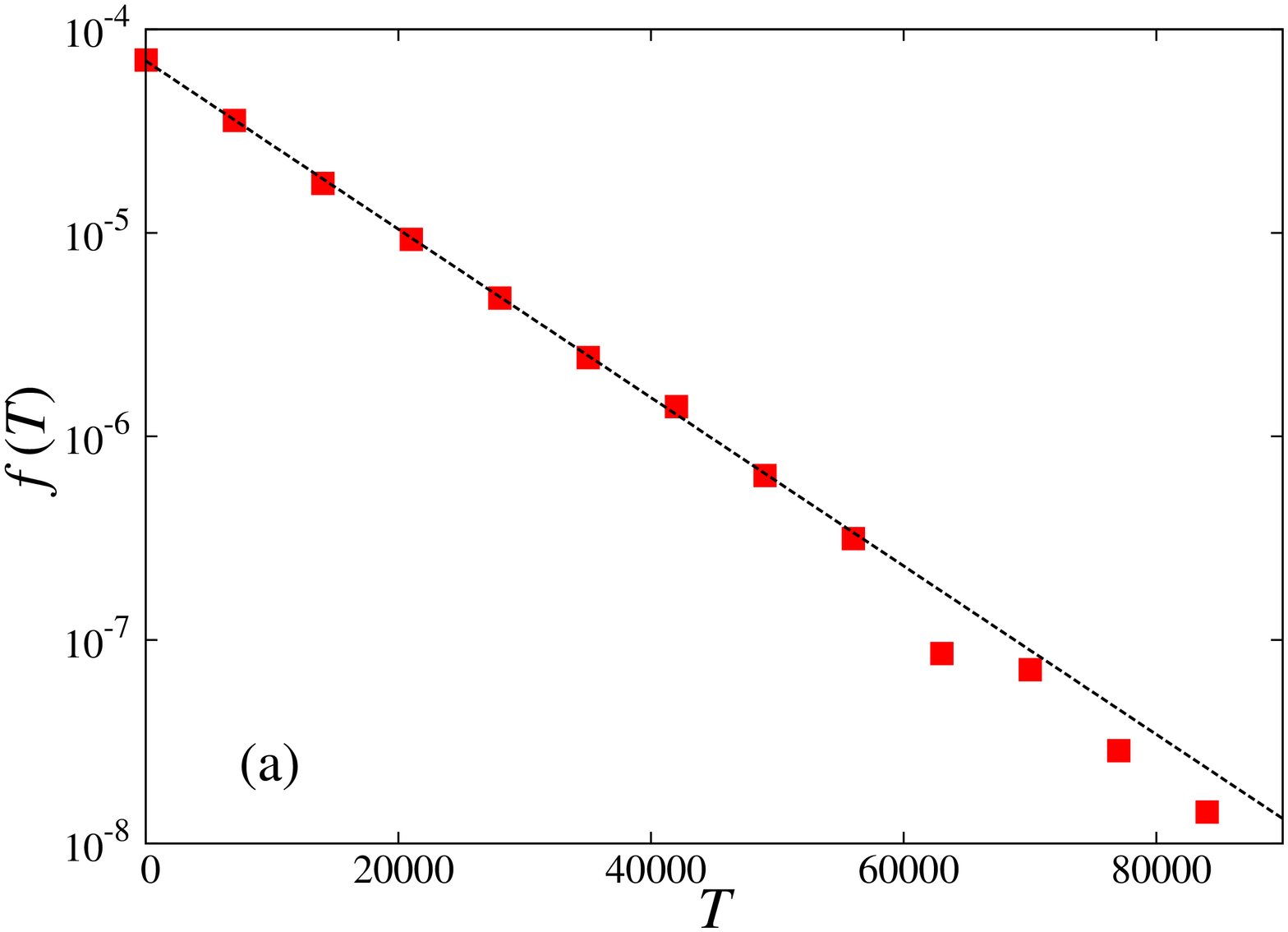}
\includegraphics[scale=0.25,angle=-90]{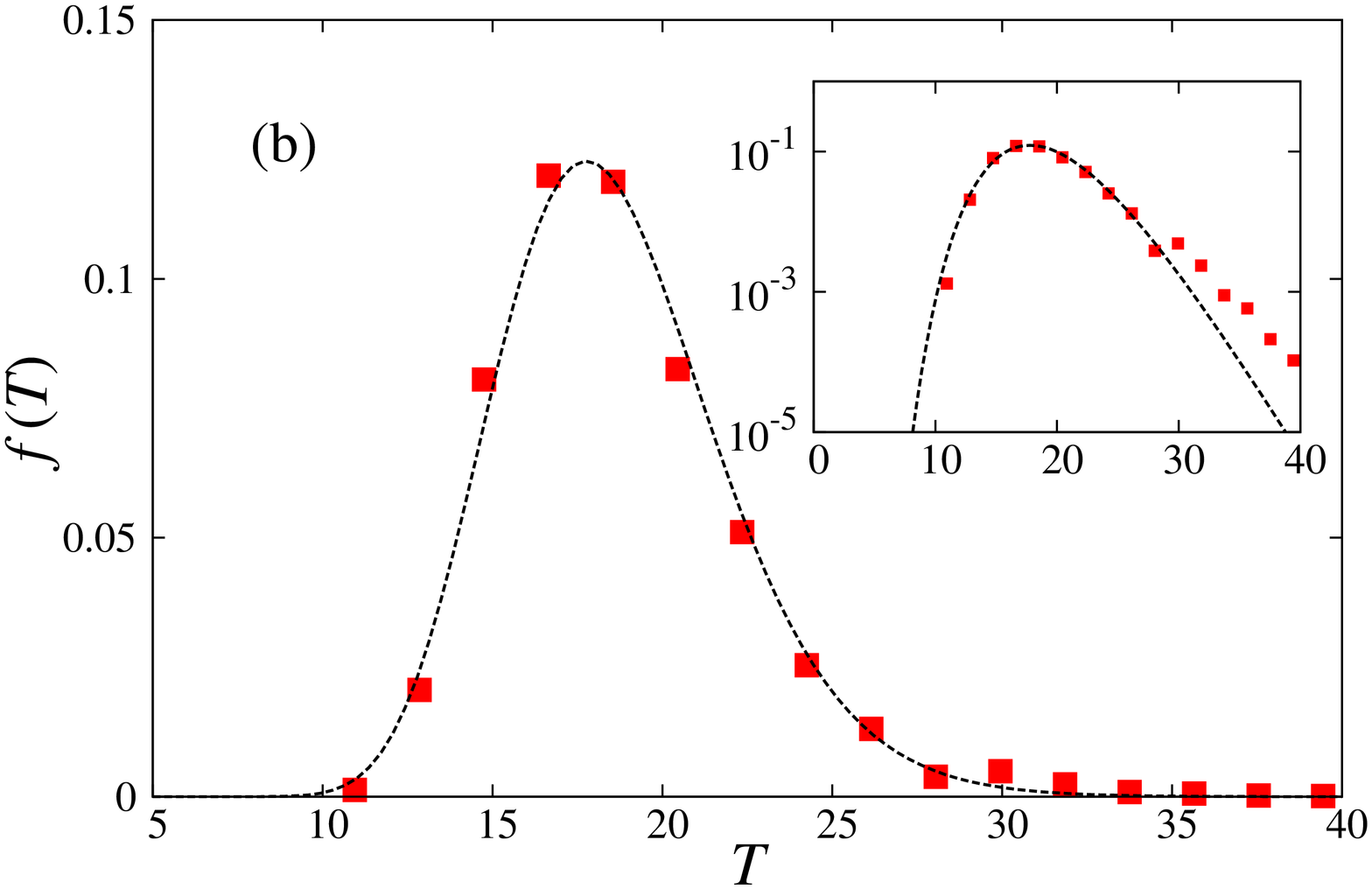}
\end{center}
\caption{Probability density distribution of the time to consensus,  for 
$(q,q_0)=(4,3)$, in which case  $\varepsilon_c = 0.25$,  
$\varepsilon = 0.30$ (a) and 0.15 (b). 
In both cases, the population size is $N=1000$. 
The symbols correspond to normalized histograms built from $10^4$ realizations. 
The dotted lines correspond to a fit of exponential (with  $\tau\simeq 10^4$) 
and log-normal (with $w\simeq 0.18$ and $x_c\simeq 18.3$) distributions,  
respectively. 
}
\label{fig:fit}
\end{figure}

Consensus time probability density functions (PDFs), estimated from the results of 
simulations, are exemplified in Fig. \ref{fig:fit}. 
For  $\varepsilon>\varepsilon_c$ [see Fig. \ref{fig:fit}(a)],  the numerical result can be well described by 
the exponential PDF $f(x)={\rm e}^{-x/ \tau}/\tau$,    
while  for $\varepsilon<\varepsilon_c$ [see Fig. \ref{fig:fit}(a)], the time to consensus can be described by 
a log-normal PDF $f(x)=  {\rm e}^{-[\ln(x/x_c)]^2/(2 w^2)}/(x\,w\,\sqrt{2\pi})$, 
where $\tau, w, x_c$ are real parameters.

In the exponential case, there is a single fitting parameter value $\tau$, 
which represents the mean, as well as the standard deviation.
In the case of the log-normal PDF, there are  two fitting parameters $w$ and $x_c$, associated 
to the mean value $x_c \,{\rm e}^{w^2/2}$ and the standard deviation 
$ (x_c)^2 {\rm e}^{w^2}({\rm e}^{w^2}-1)$.
A standard nonlinear curve fitting procedure  was used. The distribution works 
well in the central part but the tail  decays more slowly.
The full symbols in  Fig.~\ref{fig:T} represent the average time obtained from fittings, 
that is $T=\tau$ and $T=x_c \,{\rm e}^{w^2/2}$ for exponential and log-normal PDFs, respectively. 
These estimates are in good agreement with the direct average of the consensus time 
recorded for 200 samples (hollow symbols).


\section{ Large lobby size limit}
\label{appendix2}

For very large $q$, we can perform the following approximations. 
The summation in Eq.~(\ref{eq:g}) can be substituted by an integral 
and the binomial factor by its Gaussian approximation.
By integrating the resulting expression from $q_0$ to $q$, we obtain 
\begin{equation}\label{eq:erf}
g(x,q,q_0) \sim \frac{1}{2} \left\{1 - {\rm erf} \left[ \frac{q_0 - q(1-x)}{[2q(1-x)x]^{1/2}}\right] \right\} \,.
\end{equation}
In turn, in the limit $q \rightarrow \infty$, 
the error function  (erf)  can  be approximated by  the sign  function (sgn), 
where ${\rm sgn}(x)=2H(x)-1$, being $H(x)$ the Heaviside step function. Then, we get
\begin{equation}\label{eq:sign}
g(x,q,q_0) \simeq \frac{1}{2} \left\{ 1-{\rm sgn} \left[\frac{q_0}{q} -(1-x) \right] \right\}~ .
\end{equation}

Computing $V(x)=-\int_0^x v(y)dy$, from Eq.~(\ref{eq:v}), we obtain 
\begin{equation}\label{eq:pot1}
V(x)= \left\{
        \begin{array}{ll}
            - \varepsilon x  \left(1 -\frac{x}{2} \right) + \frac{x^2}{2} ~ , & x \in [0,1- \frac{q_0}{q}], \\
            \frac{(1-\varepsilon)}{2} \left( 1-\frac{q_{0}}{q} \right)^2 -
						 \varepsilon x( 1 - x)  ~ , & x \in [1- \frac{q_0}{q},\frac{q_0}{q}], \\
	    \frac{(1-\varepsilon)}{2} - x\left(1 -\frac{x}{2} \right) + \varepsilon \frac{x^2}{2} ~ , & x \in [\frac{q_0}{q},1],
        \end{array}
    \right.
\end{equation}
for  $\frac{q_0}{q} \in [\frac{1}{2},1] $, and
\begin{equation}\label{eq:pot2}
V(x) = \left\{
        \begin{array}{ll}
            - \varepsilon x  
						\left(1 -\frac{x}{2} \right) + \frac{x^2}{2} ~ , & x \in [0,\frac{q_0}{q}], \\
             (1-\varepsilon) \frac{q_{0}}{q} \left( 1- \frac{q_0}{2q} \right) - 
						  x( 1 - x)  ~ , & x \in [\frac{q_0}{q},1 - \frac{q_0}{q}] ,\\
	    \frac{(1-\varepsilon)}{2} - x\left(1 -\frac{x}{2} \right) + \varepsilon \frac{x^2}{2} ~ , & x \in [1 - \frac{q_0}{q},1],
        \end{array}
    \right.
\end{equation}
when $\frac{q_0}{q} \in [0,\frac{1}{2}] $.

Analyzing the extrema of  $V(x)$, 
we can delimit the regions in the phase diagram. 
Taking $dV/dx = 0$ in Eqs.~(\ref{eq:pot1}) and (\ref{eq:pot2}), 
we determine that there are at most three local minima localized 
at $x= \varepsilon/(1+\varepsilon),1/2,1/(1+\varepsilon)$, 
if these values belong to the corresponding intervals. 
The fact that in the large $q$ limit we obtain at most three minima suggests that this is the 
maximal number of minima in the interval $x\in [0,1]$,  for any finite $q$, despite the potential 
can be a higher order polynomial. 
Moreover, for $q=50$, sweeping $\varepsilon$, for all $q_0$,   we did not detect any 
region with more minima of the potential.  

In the lower panel of  Fig. \ref{fig:diagrams},  we present the phase diagram  in this  limit. 
Inside the OD region (yellow), where the potential has three minima,   
the dotted line represents the points where the potential 
$V$ has the same value at the three minima. 
Above the dotted line, the central minimum becomes deeper.

The boundary between  OD and D  phases is described by
\begin{equation}\label{eq:bound}
\varepsilon_c = \left\{
        \begin{array}{cl}
          q_0/(q-q_0)  \,, & \quad  0 \leq q_0/q \leq 1/2 \,,\\
            q/q_0 -1    \,, & \quad 1/2 \leq  q_0/q \leq 1 \,,\\       
	    \end{array}
    \right.
\end{equation}
obtained by the condition that the lateral minima are at the interior borders 
of the corresponding intervals. 

The O phase became restricted to the central vertical segment at $x=1/2$.

Differently, in the $q$-voter case, for very large $q$, 
only disorder is possible, except 
when $\varepsilon=0$, in which case one observes the same behavior that in the 
voter model.


{\bf Acknowledgments:} 
We acknowledge Brazilian agencies CNPq and FAPERJ for partial financial support.

{\bf Note:}
After publication, we became aware of Ref. ~\cite{threshold-sznajd}, 
where a threshold is also introduced in the $q$-voter model. 
However, these results do not overlap with ours since, besides 
different approaches are used, the model of Ref.~\cite{threshold-sznajd} 
sets $\varepsilon=0$, including independence and anti-conformity instead.

\end{document}